\documentclass[%
reprint,
superscriptaddress,
amsmath,amssymb,
aps,
floatfix,
twocolumn
]{revtex4}
\usepackage{bm}
\usepackage{graphicx}% Include figure files
\usepackage{dcolumn}% Align table columns on decimal point
\usepackage{bm}% bold math
\usepackage{xspace}
\usepackage[]{hyperref}
  \hypersetup{      
  unicode=true,         
  pdftoolbar=true,     
  pdfmenubar=true,     
  pdffitwindow=true,    
  pdfstartview={FitH},    
  pdfsubject={Sterilenus@DUNE},  
  pdfnewwindow=true,     
  pdfcreator={RevTeX},
  colorlinks=true,     
  linkcolor=blue,   
  citecolor=blue,    
  filecolor=black,   
  urlcolor=blue,           
  }
\usepackage{placeins}
\usepackage{diagbox}

\begin{document}

\preprint{APS/123-QED}

\title{Growth of Gravitational Wave Spectrum from Sound Waves in a Universe with Generic Expansion Rate}% Force line breaks with \\
\author{Yang Xiao}
\email[]{xiaoyang@itp.ac.cn}
\affiliation{Center for Theoretical Physics, Henan Normal University, Xinxiang 453007, P. R. China}

\author{Huai-Ke Guo}
\email[Corresponding author, ]{guohuaike@ucas.ac.cn}
\affiliation{ International Centre for Theoretical Physics Asia-Pacific, University of Chinese Academy of Sciences, Beijing 100190, P. R. China}

\author{Jia-Hang Hu}
\email[]{hujiahang20@mails.ucas.ac.cn}
\affiliation{ International Centre for Theoretical Physics Asia-Pacific, University of Chinese Academy of Sciences, Beijing 100190, P. R. China}

\author{Jin Min Yang}
\email[]{jmyang@itp.ac.cn}
\affiliation{
Institute of Theoretical Physics, Chinese Academy of Sciences, Beijing 100190, P. R. China}
\affiliation{Center for Theoretical Physics, Henan Normal University, Xinxiang 453007, P. R. China}

\author{Yang Zhang}
\email[Corresponding author, ]{zhangyangphy@zzu.edu.cn}
\affiliation{Center for Theoretical Physics, Henan Normal University, Xinxiang 453007, P. R. China}

\date{\today}% It is always \today, today,
             %  but any date may be explicitly specified

\begin{abstract}
% We derived here the factor $\Upsilon$, which quantifies how the gravitational wave spectrum generated by sound waves in the radiation sector grows over time, in 
% a universe with a generic expanding rate set by another dominant energy content. When the dominant energy density satisfies $\rho \propto a^{-3(1+w)}$, we found that $\Upsilon$ has a compact analytical expression: $\Upsilon =\frac{2[1-y^{3(w-1)/2}]}{3(1-w)}$, where $y = a(t)/a(t_s)$ is the ratio of the scale factor at a later time $t$ to that at $t_s$ when gravitational wave production from sound waves starts.  This generic result reduces to that derived previously for radiation-dominated and matter-dominated cases, thus generalizing previous formulas to more general cosmological contexts and providing more accurate results. The derivation relies solely on a stationary source, implying that
% this generic result of $\Upsilon$ serves as a universal factor in describing the growth of the gravitational wave production and can appear beyond cosmological phase transitions.
We derive a compact analytical expression for the growth factor $\Upsilon$, which characterizes how the gravitational wave spectrum sourced by sound waves evolves in a universe with a generic expansion history. Assuming the dominant energy density scales as $\rho \propto a^{-3(1+w)}$, we obtain $\Upsilon =\frac{2[1-y^{3(w-1)/2}]}{3(1-w)}$, where $y = a(t)/a(t_s)$ is the ratio of the scale factor at a later time $t$ to that at $t_s$ when gravitational wave production from sound waves starts. This general result reduces to known forms in radiation- and matter-dominated eras, thereby extending previous formulas to a broader class of cosmological backgrounds. The derivation assumes only that the source is stationary, making $\Upsilon$ a universal factor that captures gravitational wave growth in various scenarios—not limited to phase transitions.
\end{abstract}

%\keywords{Suggested keywords}%Use showkeys class option if keyword
                              %display desired
\maketitle

%\tableofcontents
%\section{Introduction}
% intro on PT GWs.
% intro on experimental searches LIGO-Virgoa-KAGRA, PTA, LISA/Taiji/Tianqin.
% importance of precise spectrum prediction
% lifetime dependence of spectrum

\textbf{\textit{Introduction}}--The stochastic background of gravitational waves from cosmological first-order phase transitions is one of the promising frontiers for discovering physics beyond the Standard Model~\cite{Athron:2023xlk, Roshan:2024qnv}. LIGO data has been used to search for gravitational waves originating from new physics at energy scales of $\mathcal{O}(10^3 \sim 10^6)$ TeV~\cite{Romero:2021kby, Badger:2022nwo}, far exceeding the energy scale accessible by the current large colliders. Additionally, pulsar timing array experiments reported evidences for a stochastic gravitational wave background, indicating QCD phase transitions or strongly supercooled phase transitions as possible sources~\cite{Xue:2021gyq, NANOGrav:2021flc, NANOGrav:2023gor, EPTA:2023fyk, EPTA:2023xxk, Xiao:2023dbb,Han:2023olf, Ahmadvand:2023lpp, Zheng:2024tib, Bian:2020urb, Bian:2023dnv}. Space-based gravitational waves detectors, such as LISA~\cite{amaro2017laser}, Taiji~\cite{10.1093/nsr/nwx116, Ruan:2018tsw} and Tianqin~\cite{TianQin:2015yph, TianQin:2020hid, Luo:2020bls}, are currently in preparation and expected to come into play in the coming decades.

To ascertain whether the detected gravitational waves indeed originated from the early universe phase transitions, a profound understanding of the relevant processes is essential. It is widely believed that phase transition gravitational waves can primarily be generated through three mechanisms: bubble wall collisions~\cite{PhysRevD.47.4372, PhysRevLett.69.2026, Jinno:2016vai, Zhao:2022cnn, Di:2020kbw, Li:2023yzq}, sound waves~\cite{Hindmarsh:2013xza, Hindmarsh:2015qta}, and magnetohydrodynamic (MHD) turbulence~\cite{Caprini:2009yp, Kahniashvili:2008pe, Kahniashvili:2009mf, Caprini:2015zlo, RoperPol:2019wvy}. Gravitational waves produced by bubble walls can be well-described by the envelope approximation, and an analytical expression has been obtained~\cite{Jinno:2016vai}. In thermal plasma phase transitions, gravitational waves induced by sound waves are dominant and have been extensively studied, while those from MHD turbulence are sub-dominant and related studies remain highly uncertain. Although the behavior of gravitational waves induced by sound waves can be reasonably described by the sound shell model, simulation results suggest the presence of additional mechanisms affecting the final spectrum~\cite{Cutting:2019zws}. This indicates a need for further investigation into this process. Recently, Ref.~\cite{Cai:2023guc} focused on the forced propagating contribution from the initial collision stage of sound shells and achieved better agreement with simulations compared to the sound shell model. Ref.~\cite{Ellis:2020awk} compared several models and discovered that the parameter regions with high signal-to-noise ratios (SNR) often coincide with short duration of sound waves, potentially reducing the observable gravitational wave spectrum. In our previous work~\cite{Guo:2020grp}, we calculated the gravitational wave spectrum of the sound shell model more carefully, and was able to extract an overall factor connected with the duration of sound waves, called $\Upsilon$. Ref.~\cite{RoperPol:2023dzg} pointed out that some assumptions in the sound shell model lead to incorrect infrared behavior and extended the $\Upsilon$ factor to the infrared regime.
However, most of the above calculations are in the radiation-dominated universe and have not been verified in other universe scenarios. If the equation of state of the universe deviates from the ultra-relativistic case during a phase transition, the gravitational wave spectrum may be modified~\cite{Giombi:2024kju}. Therefore, it is both necessary and interesting to extend these results to a more general cosmological period to reproduce the potential gravitational wave generation processes as accurately as possible.

The aim of this Letter is to follow our previous work and try to extend the $\Upsilon$ factor from the radiation-dominated universe and the matter-dominated universe to an arbitrary single-component dominated universe. We outline the main assumptions and derivation below, with technical details provided in the Supplemental Appendix.

\textbf{\textit{Conventions}}-We first consider the gravitational waves in the FLRW universe, described by the metric
\begin{equation}
    {\rm d}s^2 = -{\rm d}t^2 + a(t)^2\left[\delta_{ij} + h_{ij}\right] {\rm d}\mathbf{x}^2,
\end{equation}
where $h_{ij}$ is the transverse traceless part of the tensor-mode perturbations. Gravitational waves are sourced by the transverse traceless part of the perturbed stress-energy tensor of the matter content $\pi_{ij}^{T}(t, \mathbf{x})$. Those two quantities are connected by the Einstein equation, which could be written in Fourier space as 
\small
\begin{equation} \label{eq: GW equation}
    h^{\prime \prime}_{ij}(t, \mathbf{q}) + 2\frac{a^{\prime}}{a}h_{ij}^{\prime}(t, \mathbf{q}) + q^2 h_{ij}(t, \mathbf{q}) = 16\pi G a^2 \pi_{ij}^{T}(t, \mathbf{q}),
\end{equation}
where $\prime$ represents a derivative with respect to conformal time $\eta$ to distinguish it from the one with respect to coordinate time, denoted by a dot. The gravitational wave energy density is defined as 
\begin{equation}
    \rho_{\rm GW} = \frac{\left< \dot{h}_{ij}(t, \mathbf{x}) \dot{h}_{ij}(t, \mathbf{x})\right>}{32 \pi G},
\end{equation}
with the angle brackets $\left< ... \right>$ representing an average over the ensemble and the whole space. Following the Ref.\cite{Hindmarsh:2019phv, Hindmarsh:2016lnk} and considering the overall spatial homogeneity of the universe, we can define the unequal time correlator (UETC) spectrum $P_{\dot{h}}$ as 
\begin{equation} \label{eq: Ph}
    \left<\dot{h}_{ij}(t, \mathbf{q}_1) \dot{h}_{ij}(t, \mathbf{q}_2)\right> = (2 \pi)^3 \delta^3(\mathbf{q}_1 + \mathbf{q}_2) P_{\dot{h}}(q_1,t).
\end{equation}
Then the gravitational wave energy density can be reorganized as 
\begin{equation}
    \rho_{\rm GW} = \frac{1}{64 \pi^3 G}\int {\rm d} q \ q^2 P_{\dot{h}}(q,t),
\end{equation}
and the gravitational wave energy density spectrum follows 
\begin{equation}
    \frac{{\rm d}\rho_{\rm GW}(t)}{{\rm d} {\rm ln} q} = \frac{q^3 P_{\dot{h}}(t,q)}{64 G \pi^3 }.
\end{equation}
In practice, the energy density is often used in the form of dimensionless fraction $\Omega_{\rm GW}(t) = \rho_{\rm GW}(t)/\rho_{\rm c}(t)$, where $\rho_{\rm c}$ is the critical energy density at time $t$. So the dimensionless version of the spectrum is 
\begin{equation}
    \mathcal{P}_{\rm GW} \equiv \frac{{\rm d}\Omega_{\rm GW}(t)}{{\rm d} {\rm  ln}q} = \frac{q^3 P_{\dot{h}}(t,q)}{24 \pi^2 H^2} = \frac{q^3 P_{h^{\prime}}(t,q)}{24\pi^2 H^2 a^2}.
\end{equation}

% The gravitational wave equation Eq.(\ref{eq: GW equation}) could be solved by the Green's function with the following boundary conditions
% \begin{align}
%    & G(\tilde{\eta} \le \tilde{\eta}_{0}) = 0 ,  \\
% & \left.\frac{\partial G(\tilde{\eta}, \tilde{\eta}_{0})}{\partial \tilde{\eta}}\right|_{\tilde{\eta} = \tilde{\eta}_0^{+}} = 1,
% \end{align}
% where $\tilde{\eta} = q \eta$ is a dimensionless quantity and the subscript zero indicates the time when the phase transition begins. The solution is then given by
% \begin{equation}
%     h_{ij}(t, \mathbf{q}) = 16 \pi G \int_{\tilde{\eta}_0}^{\tilde{\eta}} {\rm d}\tilde{\eta}^{\prime} G(\tilde{\eta}, \tilde{\eta}^{\prime}) \frac{a^2(\tilde{\eta}^{\prime}) \pi_{ij}^{T}(\tilde{\eta}^{\prime}, q)}{q^2}.
% \end{equation}
% Inserting the above solution into the left hand of Eq.(\ref{eq: Ph}), we could obtain

The gravitational wave equation Eq.~(\ref{eq: GW equation}) can be solved using the Green's function method, allowing the spectrum function $P_{h'}$ to be expressed as
\small
\begin{align} \label{eq: Ph_prime}
    P_{h^{\prime}}(q_1, \eta) &=(16 \pi G)^2 \int_{\tilde{\eta}_0}^{\tilde{\eta}}{\rm d}\tilde{\eta}_1  \int_{\tilde{\eta}_0}^{\tilde{\eta}} {\rm d}\tilde{\eta}_2 
    \left\{ \frac{\partial G(\tilde{\eta}, \tilde{\eta}_1)}{\partial \tilde{\eta}}  
    \frac{\partial G(\tilde{\eta}, \tilde{\eta}_2)}{\partial \tilde{\eta}} \right. \notag\\
    &\left. ~~~~~~~~~~~~~~~~\times \frac{a^2(\eta_1) a^2(\eta_2)}{q^2} \Pi^2(q_1, \eta_1, \eta_2) \right\} , 
\end{align}
\normalsize 
where we have used the UETC of $\pi_{ij}^{T}$
\small
\begin{equation}
    \left<\pi_{ij}^{T}(\eta_1, \mathbf{q}_1) \pi_{ij}^{T}(\eta_2, \mathbf{q}_2)\right> = (2 \pi)^3 \delta^3(\mathbf{q}_1 + \mathbf{q}_2)\Pi^2(q_1, \eta_1, \eta_2),
\end{equation}
$G$ is the Green's function and $\tilde{\eta} = q \eta$ is dimensionless conformal time.
\normalsize 
For sound waves, the transverse-traceless part of the energy-momentum tensor, $\pi_{ij}^{T}$, is given by
%In sound shell model, the gravitation wave comes from the first-order linear approximation of fluid motion, which means
\begin{equation}
    \pi_{ij}^{T} =\Lambda_{ij, kl}\left(\bar{p} + \bar{e}\right)v^k v^l,
\end{equation}
where $\Lambda_{ij, kl}$ is the transverse-traceless projection operator, $\bar{p}$ and $\bar{e}$ are the averaged pressure and energy density. In first-order phase transition, the fluid velocity $v_i$ in the vicinity of an individual vacuum bubble can be obtained from hydrodynamic. The full velocity field in space can then be modeled as a linear superposition of these individual velocity profiles. Under this case, correlation function $\Pi$ becomes approximately stationary, depending only on the time difference $\eta_1 - \eta_2$. This leads to a factorized form of the gravitational wave spectrum

In order to explicitly display the dependence on the scale factor, we define the rescaled and dimensionless quantity $\tilde{\Pi}$ as \cite{Guo:2020grp}
\small
\begin{equation}
    \Pi^2(q, \eta_1, \eta_2) = \frac{a_s^8}{a^4(\eta_1) a^4(\eta_2)} \left[\left(\bar{\tilde{e}} + \bar{\tilde{p}}\right)\bar{U}_f^2\right]^2 L_f^3 \tilde{\Pi}^2(qL_f, q\eta_1, q\eta_2),
\end{equation}
\normalsize
where $a_{s}$ is the scale factor when the source begins active, $\bar{\tilde{e}}$ and $\bar{\tilde{p}}$ are the rescaled averaged energy density and pressure, $\bar{U}_{f}$ is the magnitude of the fluid velocity and $L_f$ is the characteristic length during the phase transition, which is usually chosen as the mean bubble separation $R_{*}$.

\textbf{\textit{$\Upsilon$~factor~in~singlet~component~universe}}-By the continuity equation for a perfect fluid
\begin{equation} 
    \dot{\rho} + \frac{3\dot{a}}{a}(\rho + P) = 0 ,
\end{equation}
and the constant equation of state between pressure and energy density $P = w \rho$, the scale factor could relate to the energy density as
\begin{equation} \label{eq: rho and scale factor}
    \rho \propto a^{-3(1+w)}.
\end{equation}
Then the different $w$ corresponds to the different single-component matter which dominates in the universe. We treat $w$ as a fixed input constant in this paper, ignoring variations in this parameter. This assumption is valid if the timescale of the gravitational wave production is relatively short.

% Inserting Eq.(\ref{eq: rho and scale factor}) into the Friedmann equation of flat universe
% \begin{align}
%     \frac{{\rm d}a}{{\rm d}\eta} = a^2\sqrt{\frac{8\pi G}{3} \rho} 
%     = a^2 \sqrt{H_s^2 \left(\frac{a}{a_s}\right) ^{-3(1+w)}},
% \end{align}
% and defining $y(\eta) = a(\eta)/a_s$, following our previous derivation \cite{Guo:2020grp}, we could obtain
% \begin{equation}
%     \frac{{\rm d} y}{{\rm d}\eta} = a_s H_s y^{(1-3w)/2},
% \end{equation}
% where $H_s$ is the Hubble constant when the source begins active. With the initial condition $y(\eta_s) = 1$, the solution of the above equation is 
% \begin{equation}
%     y^{\frac{1+3w}{2}} = \frac{3w+1}{2}(\eta-\eta_s)H_sa_s + 1.
% \end{equation}
% Changing the variable from $\eta$ to $y$, the gravitational wave equation Eq.(\ref{eq: GW equation}) could be rewritten as 

Inserting Eq.~(\ref{eq: rho and scale factor}) into the Friedmann equation for a flat universe and introducing a rescaled time variable $y(\eta) \equiv a(\eta)/a_s$, where the subscript $s$ denotes the moment when the source becomes active, Eq.~(\ref{eq: GW equation}) can be recast in the following form
\small
\begin{equation}
    \frac{{\rm d}^2 h_{ij}}{{\rm d} y^2} + \frac{{\rm d}h_{ij}}{{\rm d}y}\left(\frac{5-3w}{2y}\right) + \frac{q^2 y^{3 w -1}}{(H_s a_s)^2}h_{ij} = \frac{16 \pi Gy^{3w-1}a^2}{(H_sa_s)^2}\pi_{ij}^{T}.
\end{equation}
\normalsize
Then denoting $q^2/(H_sa_s)^2 = \tilde{q}^2$ and $\tilde{y} = y \tilde{q}^n$, the equation could be simplified further
\small
\begin{equation}
    \tilde{y}^{1-3w}\frac{{\rm d}^2 h_{ij}}{{\rm d} \tilde{y}^2} + \tilde{y}^{-3 w}\frac{{\rm d} h_{ij}}{{\rm d} \tilde{y}}(\frac{5-3w}{2 }) + h_{ij} = \frac{16 \pi G a^2}{q^2} \pi_{ij}^{T},
\end{equation}
\normalsize
by setting $2n = 2 - n(3w-1)$. The corresponding Green's function is the homogeneous solution of this equation and could be expressed as \cite{bowman2012introduction}
\begin{align} \label{eq: Green function}
    G(\tilde{y}, \tilde{y}_0) &= \tilde{y}^{\frac{3w-3}{4}}\left[C_1J_{\nu}\left(\frac{2}{3w+1}\tilde{y}^{\frac{3w+1}{2}}\right)  \right. \notag \\
    &\left.+ C_2J_{-\nu}\left(\frac{2}{3w+1}\tilde{y}^{\frac{3w+1}{2}}\right)\right],
\end{align}
where $J_{\nu}(x)$ is the Bessel function of the first kind and the $\nu = (3w-3)/(6w+2)$ should not be an integer. If $\nu $ is an integer, the $J_{-\nu}$ in the above expression should be replaced with the Bessel function of the second kind to satisfy the condition of linear independence \cite{bowman2012introduction}. In this paper, we confine ourselves to values of $w$ that satisfy the strong-energy condition, i.e., $1+3w > 0$, which implies that almost all values of $\nu$ are not integers. 

% The constants $C_1$ and $C_2$ can be determined by the following initial conditions
% \begin{align}
%     &G(\tilde{y} \le \tilde{y}_0) = 0, \\
%     &\left.\frac{\partial G(\tilde{y}, \tilde{y}_0)}{\partial \tilde{y}}\right|_{\tilde{y} = \tilde{y}_{0^{+}}} = \tilde{y}_{0}^{3 w + 1},
% \end{align}
% with the following results:
% \begin{align}
%     C_1 &= -\frac{b}{b_1 a - b a_1} \tilde{y}_{0}^{\frac{3w+1}{4}},\\
%     C_2 &= \frac{a}{b_1 a - b a_1} \tilde{y}_{0}^{\frac{3w+1}{4}},
% \end{align}
% where $a = J_{\nu}\left(\frac{2}{3w+1}\tilde{y}_0^{\frac{3 w + 1}{2}}\right)$, $b = J_{-\nu}\left(\frac{2}{3w+1}\tilde{y}_0^{\frac{3 w + 1}{2}}\right)$ and $a_1$ ($b_1$) is the corresponding derivative with respect to its variable, ${\rm d}J_{\nu}(z)/{\rm d}z$ (${\rm d}J_{-\nu}(z)/{\rm d}z$).

With the Green's function, the dimensionless gravitational wave spectrum $\mathcal{P}_{\rm GW}$ is given by
\small
\begin{align} \label{eq: Pgw}
    &\mathcal{P}_{\mathrm{GW}}\left(y, q R_{*}\right)=\frac{\left[16 \pi G \left(\bar{\tilde{e}} + \bar{\tilde{p}}\right) \bar{U}_f^2\right]^2 \left(qR_{*}\right)^3}{24 \pi^2 H^2H_s^2 y^{3w+1}}\notag \\
    &\times \int_{y_s}^{y} {\rm d} y_1 \int_{y_s}^{y} {\rm d} y_2 
    \left\{ \left(\frac{q}{a_sH_s}\right)^{4n-2}\frac{\partial G\left(\tilde{y}, \tilde{y}_1\right)}{\partial \tilde{y}} \frac{\partial G\left(\tilde{y}, \tilde{y}_2\right)}{\partial \tilde{y}} \right. \notag \\ 
    &\left. ~~~~~~~~~~~~~~~~~~~~~~~~\times \frac{1}{y_1^{2} y_2^{2}} \frac{\tilde{\Pi}^2\left(q R_{*}, q |\eta_1- \eta_2| \right)}{\tilde{q}^2} \right\} .
\end{align}
\normalsize 
Because the source is largely stationary, the dependence of $\mathcal{P}_{\rm GW}$ on the conformal time is only through the difference
\begin{equation} \label{eq: eta_diff}
    q|\eta_1 - \eta_2| = \frac{2qR_{*}}{R_{*}a_sH_s(3w+1)}\left|y_1^{\frac{3w+1}{2}} - y_2^{\frac{3w+1}{2}}\right|.
\end{equation}
This results in a factorized form of the spectrum $\mathcal{P}_{\rm GW}$ for most frequencies in the range near the peak~\cite{Guo:2020grp}. This factor, called $\Upsilon$ in previous studies, can be obtained in the
following way.

Firstly we change the integral variables from $y_1$ and $y_2$ to $y_{+}$ and $y_{-}$
\begin{align}
y_{-} &= \frac{2}{3 w +1} (y_1^{\frac{3w+1}{2}} - y_2^{\frac{3 w +1}{2}}),  \\
y_{+} &= \frac{1}{3 w +1} (y_1^{\frac{3w+1}{2}} + y_2^{\frac{3 w +1}{2}}). 
\end{align}
% The corresponding Jacobian determinant is $(y_1y_2)^{\frac{1-3w}{2}}$ and the new limits of integration are
% \begin{align}
%     &0 \le y_{-} \le (y^{\frac{3 w +1}{2}} -1) \frac{2}{3 w + 1}, \\ 
%     &\frac{2}{3 w + 1} + \frac{y_{-}}{2} \le y_{+} \le -\frac{y_{-}}{2} + \frac{2}{3w+1} y^{\frac{3 w +1}{2}},
% \end{align}
% and
% \begin{align}
%   &  (1 - y^{\frac{3 w +1}{2}}) \frac{2}{3 w +1} \le y_{-} \le 0, \\
%   &  -\frac{y_{-}}{2} + \frac{2}{3w+1} \le y_{+} \le \frac{y_{-}}{2} + y^{\frac{3w+1}{2}} \frac{2}{3 w + 1}.
% \end{align}
In additional, to explicitly show the $a^{-4}$ behavior of the gravitational radiation, we also denote
\begin{equation}
    \frac{\partial G(\tilde{y}, \tilde{y}_1)}{\partial \tilde{y}} \frac{\partial G(\tilde{y}, \tilde{y_2})}{\partial \tilde{y}} = \frac{\mathcal{G}(\tilde{y}, \tilde{y}_1, \tilde{y}_2)}{\tilde{y}^{3-3w}}.
\end{equation}
As a result, we could further simplify Eq.(\ref{eq: Pgw}) as 
\begin{align} 
    &\mathcal{P}_{\mathrm{GW}}\left(y, q R_{*}\right)=\frac{\left[16 \pi G\left(\bar{\tilde{e}} + \bar{\tilde{p}}\right) \bar{U}_f^2\right]^2 (qR_{*})^3}{24 \pi^2 H^2H_s^2 y^{4}} \notag \\  
    & ~~~~\times \int {\rm d} y_{-} \frac{\tilde{\Pi}^2\left(q |\eta_1- \eta_2| \right)}{\tilde{q}^2}    
     \int {\rm d} y_{+} \frac{\mathcal{G}\left(\tilde{y}, \tilde{y}_1, \tilde{y}_2\right)}{y_1^{\frac{3+3w}{2}} y_2^{\frac{3+3w}{2}}}.
\end{align}

It turns out that $y_{-} \ll y_{+}$ generally holds except in some special parameter spaces for sound shell model. This could be verified by noting that $\beta R_{*} = (8\pi)^{1/3}v_w \approx 3v_w$ and $R_{*}a_sH_s \sim \mathcal{O}(10^{-3})$, where $\beta$ is the value of the derivative of the logarithm of the probability of bubble nucleation with respect to time at some reference temperature \cite{Guo:2020grp}. Thus $y_{-} \sim \mathcal{O}(10^{-3})/v_w \times \beta(\eta_1  - \eta_2)$ according to Eq.(\ref{eq: eta_diff}). When $|\eta_1 - \eta_2|$ exceeds $\mathcal{O}(10)$, the $\tilde{\Pi}$ quickly decays to zero and so the effective integral interval of $y_{-}$ is very close to zero~\cite{Guo:2020grp}. On the contrary, $y_{+} \ge 1$ for the minimum of $y_1$ or $y_2$ is 1. Thus, we have $y_{+} \gg y_{-}$, which implies that we could keep the leading order of $y_{-}$ when dealing with the integration of $y_{+}$. The $\Upsilon$ factor is exactly the result of the integration of $y_{+}$ and it serves as a multiplicative factor in the expression of gravitational wave spectrum.

To calculate this complex integration, it is worth noting that the variable of the Green's function Eq.(\ref{eq: Green function}) 
is the combination $\tilde{q} = q/H_sa_s = qR_{*}/H_sa_sR_{*} \sim \mathcal{O}(10^{3}) qR_{*}$. This indicates that we could use the asymptotic expansion of the Bessel functions
\begin{align}
    J_{\nu}(x) &\sim \frac{\sqrt{\frac{\pi}{2}}{{\rm cos}\left[x - \frac{1}{4}(1 + 2 \nu)\right]}}{\sqrt{x}} ,
%    &+\frac{(1-4\nu^2){\rm sin}\left[x - \frac{\pi}{4}(1+2\nu)\right]}{4\sqrt{2\pi}x^{3/2}}
\end{align}
if $\tilde{q}$ is large. After simplification (see Appendix for detail), the integral with respect to 
$y_{+}$ can ultimately be written as
\begin{align}
     \mathcal{G}_2 &\equiv \tilde{q}^2 \frac{{\rm cos}(\tilde{q}y_{-})}{2} \Upsilon^{\nu \notin \mathbb{N}} (y;w),
\end{align}
where 
\begin{equation}
    \Upsilon^{\nu \notin \mathbb{N}}  =\frac{2\left[y^{3(w-1)/2}-1\right]}{3(w-1)}.
\end{equation}

\begin{figure}[htbp!]
\centering 
\includegraphics[width=0.48\textwidth]{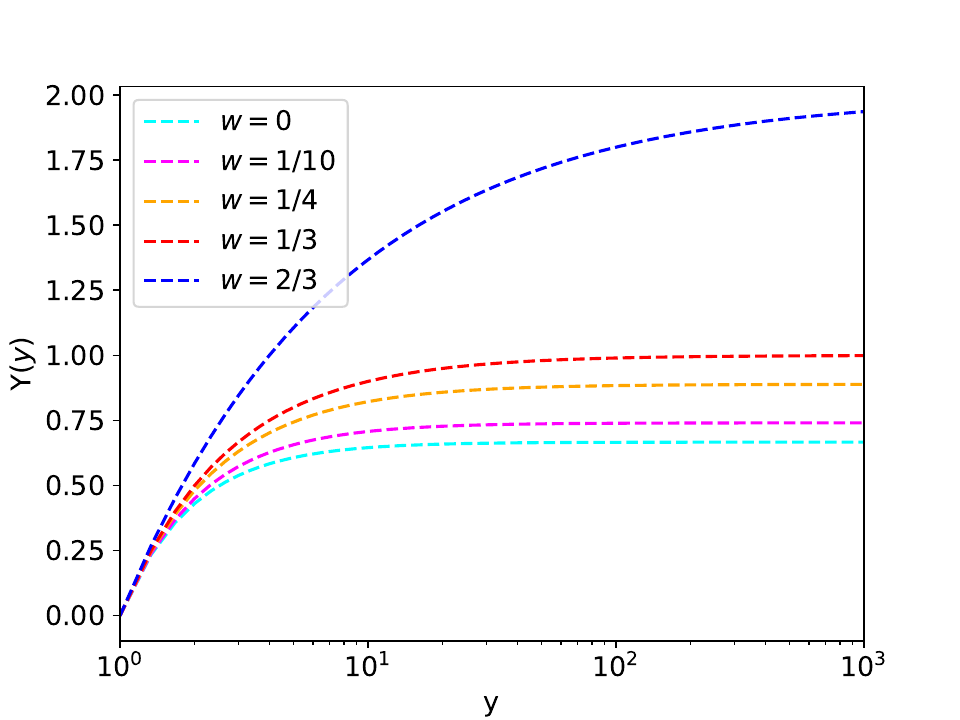}
\caption{The $\Upsilon$ factor for different $w$. In the plot, $c_k$ is set to 0.00079, and $qR_{*}$ is set to 10.}
\label{fig:Upsilon}
\end{figure}

For the case where $\nu$ is an integer, we can follow a similar simplification method to obtain the corresponding $\Upsilon$, which yields the same result as it in the non-integer case. Thus, our final expression for $\Upsilon$ is
\begin{equation}
    \Upsilon =\frac{2\left[y^{3(w-1)/2}-1\right]}{3(w-1)}.
\end{equation}
With this result, the final dimensionless gravitational wave spectrum is 
\small
\begin{align} \label{eq: Pgw_upsilon}
    \mathcal{P}_{\rm GW} = \Upsilon(y;w) \frac{\left[16 \pi G \left(\bar{\tilde{e}} + \bar{\tilde{p}}\right) \bar{U}_f^2\right]^2 \left(qR_{*}\right)^3}{48 \pi^2 H^2H_s^2 y^{4}} \int {\rm d}y_{-} {\rm cos}(\tilde{q}y_{-})\tilde{\Pi}^2.
\end{align}
\normalsize
Compared with the result of \cite{Hindmarsh:2019phv}, the $\Upsilon$ factor is an extra factor originated from the finite duration time of sound waves. The $\Upsilon$ factor for different values of $w$ is shown in Fig.{(\ref{fig:Upsilon})}. Considering the radiation-dominated universe case ($w = \frac{1}{3}$)
% \begin{align}
%     \left(P_1^2 +P_3^2-2P_1P_3{\rm cos}[\frac{3\pi(w-1)}{6w+2}]\right) = 4 \notag, \\
% \left({\rm cos}\left[\frac{3w \pi-\pi}{3w+1}\right]-{\rm cos}\left[\frac{2\pi}{3w+1}\right]\right)^2 = 4,
% \end{align}
and the matter-dominated universe case ($w = 0$),
% \begin{align}
%     \left(P_1^2 +P_3^2-2P_1P_3{\rm cos}[\frac{3\pi(w-1)}{6w+2}]\right) = 4, \notag \\
% \left({\rm cos}\left[\frac{3 w \pi-\pi}{3w+1}\right]-{\rm cos}\left[\frac{2\pi}{3w+1}\right]\right)^2 = 4,
% \end{align}
our new $\Upsilon$ factor could correctly reproduce our previous results \cite{Guo:2020grp}
\begin{align}
    \Upsilon_{\rm RD} & = 1 - \frac{1}{y} ,  \\
    \Upsilon_{\rm MD} &= \frac{2}{3}\left(1 - \frac{1}{y^{3/2}}\right).
\end{align}
% For other values of $w$, $\Upsilon$ may depend on the dimensionless combination $kR_{*}$, which differs from the two cases discussed above. This dependence on the wave number primarily influences the asymptotic value of the $\Upsilon$ factor, but does not affect its asymptotic behavior.

In our previous work, we referred to $\Upsilon$ as the suppression factor because its values are always lower than the corresponding asymptotic value. When extending to an arbitrary single-component universe, $\Upsilon$ can still act as a suppression factor for most of $w$, and the effect becomes even more pronounced when $w$ is sufficiently large, as shown by the blue line in Fig.(\ref{fig:Upsilon}). When the universe is dominated by the scalar field and some other matters, it is possible that the corresponding effective $w$ is about 2/3 and its $\Upsilon$ could be larger than one, even if the duration time is not very long. So it would indeed hold significant meaning if the source of gravitational waves went through some kinds of non-standard cosmological processes. In addition, this extension also makes strides in precisely calculating the gravitational wave spectrum during the transition period, when the primary constituents of the universe undergo changes. 

An interesting aspect is that our derivation relies only on the conditions that the source is stationary and that the characteristic quantity $R_{*}a_sH_s$ is relatively small. Consequently, the resulting $\Upsilon$ factor is actually independent of the sound shell model. This suggests that $\Upsilon$ may serve as a more universal factor to describe how gravitational waves induced by a stationary source evolve over time \cite{RoperPol:2023dzg}.

\begin{figure}[htbp!]
\centering 
\includegraphics[width=0.48\textwidth]{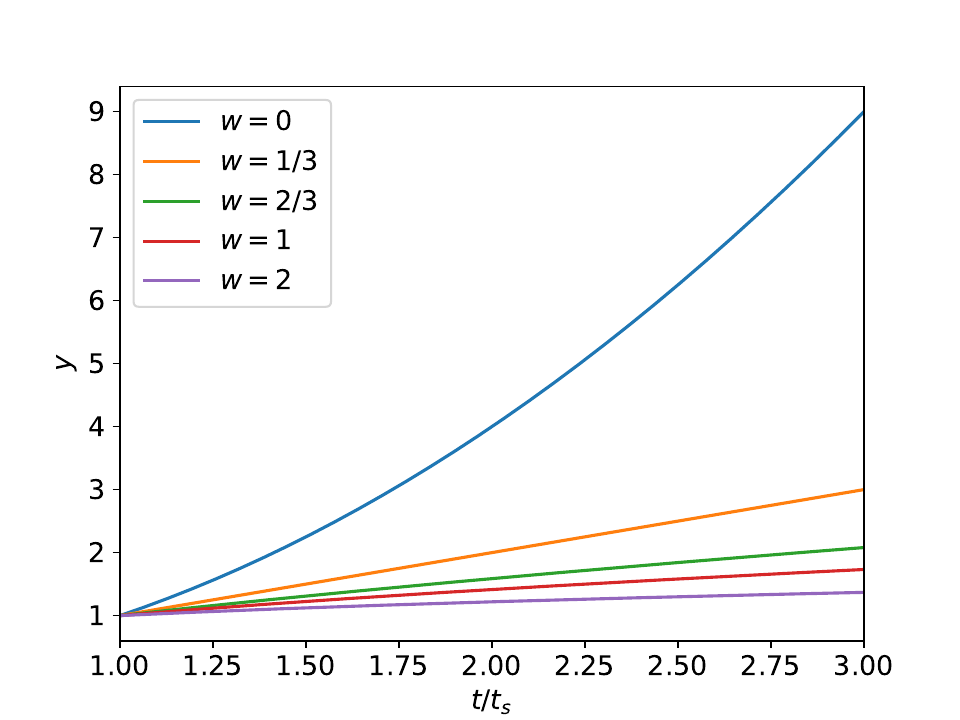}
\caption{Schematic diagram of $y$ as a function of time $t$.}
\label{fig:y(t)}
\end{figure}

It should be noted that our new $\Upsilon$ is ill-defined when $w = 1$. In order to investigate the behavior of the $\Upsilon$ in this vicinity, we perform a Taylor expansion as below 
\begin{align} \label{eq: upsilon theta}
     \left.\Upsilon(y;w)\right|_{w \to 1} &\approx {\rm ln}(y) + \mathcal{O}(w-1).
\end{align}
We find that although $\Upsilon$ does not diverge in the vicinity of $w=1$, it behaves like a logarithmic function, suggesting that $\Upsilon$ tends to infinity when the active time of the source is sufficiently long. 
To understand the underlying reasons, we reexamine Eq. (\ref{eq: G2}) and observe that the logarithmic behavior stems from the power of $y_{+}$ in the integrand, which equals $-1$ when $w=1$. This prompts us to analyze the composition of this power. There are three contributions to this power: (1) the Jacobian determinant provides a factor of $(y_1 y_2)^{\frac{1-3w}{2}}$; (2) the Green's function contributes a factor of $y_{+}^2$; (3) the UETC of the energy-momentum tensor $\Pi^2$ introduces a factor of $y_1^4 y_2^4$, which is canceled by $a^2(\eta_1) a^2(\eta_2)$, leaving a term $y_1^2 y_2^2$, as shown in Eq.(\ref{eq: Ph_prime}). Thus, we can express $\Upsilon$ roughly as
\begin{align} \label{eq: explain_lny}
    \Upsilon &\sim \int d y_{+} ~ y_{+}^2
\times \frac{\left(y_1 y_2\right)^{\frac{1-3w}{2}}}{y_1^{2} y_2^{2}} \notag \\
&= \int d y_{+} \frac{{\rm Green's~function}}{{\rm Source}} \times {\rm Jacobian~Det} \notag\\
&= \int d y_{+} ~ \frac{y_{+}^{2}}{y_{+}^{\frac{8}{1+3w}}} y_{+}^{\frac{2-6w}{1+3w}}= \int d y_{+} ~y_{+}^{\frac{-4}{1+3w}} ,
\end{align}
where we have assumed $y_{-}$ to be close to zero and so $y_1 \approx y_2 \approx y_{+}^{\frac{2}{3w+1}}$. In the above equation, the denominator represents the generation of gravitational waves induced by an instant source, the numerator reflects the source's dilution effect, and the determinant accounts for the impact of the coordinate transformation. It is clear that the first two terms compete and cancel each other when $w = 1$, leaving only the determinant to contribute a factor of $1/y_{+}$ and resulting in logarithmic behavior. This fact inspires us to revisit our coordinate system to obtain a qualitative explanation. For $w \neq -1$, scale factor $a$ is proportional to $t^{\frac{2}{1+3w}}$, and so 
\begin{equation}
    y = \frac{a(t)}{a_s} = \left(\frac{t}{t_s}\right)^{\frac{2}{3(1+w)}}.
\end{equation}
If we fix the value of $y$, then larger values of $w$ will result in a longer duration for the slowly increasing behavior, as seen in Fig.(\ref{fig:y(t)}). This is equivalent to the gravitational wave source dilutes more slowly, allowing more time for the production of gravitational waves, which causes the asymptotic value of $\Upsilon$ to increase with $w$, and ultimately results in logarithmic behavior when $w = 1$.

One might ask how to deal with the infinity of $\Upsilon$ when $y \to +\infty$. In fact, we have implicitly assumed in the above procedure that when the gravitational wave source is active, the equation governing the background energy density remains unchanged. If the gravitational wave source lasts long enough, the dominant component of the cosmic background may change, which means that $w$ is actually a function of $y$. Therefore, we cannot simply take $y$ to infinity while fixing $w$ as a constant. 

In addition, the universe is in the so-called kination period when $w = 1$, during which the kinetic energy of a free scalar field dominates. Ref.\cite{Gouttenoire:2021jhk} demonstrated that this period must occur either immediately after inflation or following the matter-dominated era. Since the energy density during the kination period decreases rapidly as $a^{-6}$, the duration of this period is strictly restricted to be consistent with the results of Big Bang Nucleosynthesis and Cosmic Microwave Background. Thus, even if $w$ is always one, $y$ still cannot be taken as positive infinity.

Lastly, $\Upsilon$ also does not have an asymptotic value when $w > 1$. However, due to the requirements of causality, this case can be disregarded. Specifically, the energy density will be less than the pressure density when $w > 1$, leading to the speed of sound exceeds the speed of light.

\textbf{\textit{Spectrum~Today}}--With $\Upsilon$ obtained, the gravitational wave spectrum today can be written in the following factorized form:
\begin{align} 
 \Omega_{\text{GW}}h^2(f)|_{\text{today}} &= \text{Redshift factor} \times \mathcal{P}_{\rm GW},
\end{align}
where $\mathcal{P}_{\rm GW}$ can be found in Eq.(\ref{eq: Pgw_upsilon}) and the redshift factor is
\begin{equation}
    {\rm Redshift~factor} = \left(\frac{a_{e}}{a_0}\right)^4 \left(\frac{H_{e}}{H_0}\right)^2 h^2,
\end{equation}
with $a_e$ and $H_e$ representing the scale factor and Hubble parameter at the moment when gravitational waves cease to be produced, $a_0$ and $H_0$ denoting the current scale factor and Hubble parameter, while $h \approx 0.678$ being the Hubble parameter today in unit of $100 {\rm km}/s/{\rm Mpc}$.

Expanding Eq.(\ref{eq: Pgw_upsilon}) and using the explicit expressions of $H$ and $H_s$, we can rewrite it as 
\begin{equation}
    \mathcal{P}_{\rm GW} = 3 \Gamma^2 \chi^2 \bar{U}_f^4 H_s a_s R_{*} \mathcal{A} \mathcal{S_{\rm SW}} \frac{y^{3(w+1)}}{y^4} \Upsilon(y;w),
\end{equation}
 where $\chi$ represents the ratio of the source energy density to the total energy density of the universe at the source active time $t_s$, $\Gamma$ is the ratio of source enthalpy density to the corresponding energy density and $A \mathcal{S_{\rm SW}}$ is the spectrum function that denotes the integration
\begin{equation}
   \mathcal{A} \mathcal{S_{\rm SW}} = \frac{\left(qR_{*}\right)^3}{4 \pi^2} \int {\rm d}y_{-} {\rm cos}(\tilde{q}y_{-})\tilde{\Pi}^2(y_{-}, qR_{*}).
\end{equation}
In principle, the spectrum function can be derived by either calculating the integration described above or utilizing the results from the numerical simulations. For radiation-dominated era, the numerical result is
\begin{equation}
\mathcal{A}\mathcal{S_{\rm SW}} = 0.008 \times \left(\frac{f}{f_{\rm  SW}}\right)^3 \left[\frac{7}{4 + 3 
 \left(f/f_{\rm SW}\right)^2}\right]^{7/2},
\end{equation}
where $f_{\rm SW}$ is the peak frequency of the corresponding gravitational wave~\cite{Hindmarsh:2017gnf}. 

In phase transition, we can further express $R_{*}$ by the inverse transition duration time $\beta$ and so the characteristic quantity $H_s a_s R_{*}$ can be rearranged as 
\begin{equation}
  a_s H_s R_{*} \approx \left(8\pi\right)^{1/3} v_w \frac{H_s}{\beta},
\end{equation} where $v_w$ is the bubble wall velocity and we have assumed that bubbles disappear quickly. With all the above ingredients, the final spectrum is
\begin{align}
    \Omega_{\text{GW}}h^2(f)|_{\text{today}} &= 4.04 \Gamma^2 \chi^2 \bar{U}_f^4  v_w \frac{H_s}{\beta} \mathcal{A}\mathcal{S}_{\rm SW} \notag \\
    &\times \left(\frac{a_{e}}{a_0}\right)^4 \left(\frac{H_{e}}{H_0}\right)^2 \frac{y^{3(w+1)}}{y^4} \Upsilon(y;w).
\end{align}

It is interesting to note the explicit dependence of the gravitational wave spectrum on $w$. Finally, we emphasize that if the duration of gravitational wave production is sufficiently long that the dominant component of the cosmic background changes, the above spectrum is no longer valid and the corresponding $\Upsilon$ must be applied for each era accordingly.

\textbf{\textit{Conclusion}}--In this work, we successfully utilized the sound shell model in an expanding universe to derive the factor $\Upsilon$, which reflects the growth of gravitational waves induced by sound waves in a single-component universe that satisfies the strong energy condition. We obtained the following observations: (1) Our new $\Upsilon$ accurately reproduces previous results for radiation-dominated and matter-dominated universes; (2) During the kination period, the new $\Upsilon$ exhibits a logarithmic behavior, driven by the competition between the rate of the gravitational wave production and the dilution rate of the gravitational wave source; (3) The new $\Upsilon$ does not depend on the sound shell model, suggesting that it may represent a more universal result and is not limited to cosmological phase transitions.

\textit{Acknowledgments}--We would like to thank Ligong Bian and Rui-Xiao Zhang for helpful discussions. This work is supported by the National Natural Science Foundation of China (12105248, 11821505, 12075300, 12335005, 12147103), by Peng-Huan-Wu Theoretical Physics Innovation Center (12047503), and by the Research Fund for Outstanding Talents from Henan Normal University (5101029470335).

\section*{Supplemental Appendix}
\subsection{Derivation of $\Upsilon$ factor}

By substituting the asymptotic form of the Bessel function, the integral over $y_{+}$ can be simplified into a fractional expression. After omitting the linear terms associated with $H_sa_sR_{*}$, the oscillatory term of its numerator $\mathcal{N}_{\rm osc}$ can be expressed as 
\begin{equation} \label{eq: num}
    \mathcal{N}_{\rm osc} = 4 \left[qR_{*}\right]^2 y^{1+3w}(\tilde{P}_1\tilde{P}_2-\tilde{P}_3\tilde{P}_4)(\tilde{P}_5\tilde{P}_6-\tilde{P}_7\tilde{P}_8),
\end{equation}
where $\tilde{P}_i$ are complex expressions involving trigonometric functions and they must be carefully handled to extract the correct results from strong oscillations. The key to significantly simplify these expressions lies in properly utilizing the fact that $c_k \equiv R_{*}a_sH_s \sim \mathcal{O}(10^{-3})$. After ignoring some small terms, we find that
\small
\begin{align}
    \tilde{P}_1 &= \tilde{P}_5= AP_1 = -2A{\rm cos}\left[\frac{3c_k\pi w-2qR_{*}y^{\frac{1+3w}{2}}}{c_k +3c_kw}\right],   \\
    \tilde{P}_2 &= B{\rm cos}\left[\frac{\pi}{4}\left(1 + \frac{6-6w}{2+6w}\right)-\frac{qR_{*}y_{+}}{c_k} + \frac{qR_{*}y_{-}}{2c_k}\right],  \\
    \tilde{P}_3 &= \tilde{P}_7 =AP_3= 2A{\rm sin}\left[\frac{c_k\pi-2qR_{*}y^{\frac{1+3w}{2}}}{c_k+3c_kw}\right], \\
    \tilde{P}_4 &= B{\rm cos}\left[\frac{\pi}{4}\left(1 + \frac{6w-6}{2+6w}\right)-\frac{qR_{*}y_{+}}{c_k} + \frac{qR_{*}y_{-}}{2c_k}\right],  \\
    \tilde{P}_6 &= B{\rm cos}\left[\frac{\pi}{4}\left(1 + \frac{6-6w}{2+6w}\right)-\frac{qR_{*}y_{+}}{c_k} - \frac{qR_{*}y_{-}}{2c_k}\right], \\
    \tilde{P}_8 &= B{\rm cos}\left[\frac{\pi}{4}\left(1 + \frac{6w-6}{2+6w}\right)-\frac{qR_{*}y_{+}}{c_k} - \frac{qR_{*}y_{-}}{2c_k}\right], 
    % A &= \frac{\sqrt{c_k+3c_kw}}{\sqrt{\pi qR_{*}y^{\frac{1+3w}{2}}}}, \notag \\
    % B &= \frac{\sqrt{2 c_k}}{\sqrt{\pi qR_{*}y_{+}}},
\end{align}
\normalsize 
where 
\begin{eqnarray}
A = \frac{\sqrt{c_k+3c_kw}}{\sqrt{\pi qR_{*}y^{\frac{1+3w}{2}}}},
\quad B = \frac{\sqrt{2 c_k}}{\sqrt{\pi qR_{*}y_{+}}}. 
\end{eqnarray} 
By denoting
\begin{align}
    x &= \frac{\pi}{4}\left(1 + \frac{6-6w}{2+6w}\right)-\frac{qR_{*}y_{+}}{c_k},  \\
    x^{\prime} &= \frac{\pi}{4}\left(1 + \frac{6w-6}{2+6w}\right)-\frac{qR_{*}y_{+}}{c_k}, \\
    z &= \frac{qR_{*}y_{-}}{2c_k},
\end{align}
Eq.(\ref{eq: num}) can be written as
\begin{align}
    \mathcal{N}_{\rm osc} & \sim A^2B^2 \left[P_1 *{\rm cos}(x+z)-P_3*{\rm cos}(x'+z)\right] \notag\\
   &\times \left[P_1*{\rm cos}(x-z)-P_3*{\rm cos}(x'-z)\right].
\end{align}
Because we want to extract the term related to $y_{-}$, we use the properties of trigonometric functions
\small
\begin{equation}
    {\rm cos}(x+z){\rm cos}(x'-z) = \frac{{\rm cos}(x+x')+{\rm cos}(x-x'+2z)}{2},
\end{equation}
\normalsize 
to obtain the desired form of $\mathcal{N}_{\rm osc}$ which includes the sum of
\small
\begin{align}
    A^2B^2 \left[P_1^2\frac{{\rm cos}(2x)}{2}-2P_1P_3\frac{{\rm cos}(x+x')}{2} + P_3^2\frac{{\rm cos}(2x')}{2}\right] ,
\end{align}
\normalsize 
and
\small
\begin{align}
    A^2B^2 &\left[P_1^2\frac{{\rm cos}(2z)}{2}-P_1P_3{\rm cos}(2z){\rm cos}(x-x')  +P_3^2\frac{{\rm cos}(2z)}{2}\right].
\end{align}
\normalsize 

The denominator $\mathcal{D}$ is the form of
\begin{equation}
    \mathcal{D}=8c_k^2 (M_1-M_2+M_3M_4) (M_5-M_6+M_7M_8),
\end{equation}
where $M_i$ are also complex expressions involving trigonometric functions. By similar steps, the $\mathcal{D}$ could be simplified as 
\begin{equation}
    \mathcal{D} = \frac{32c_k^4}{\pi^2} \frac{\left({\rm cos}\left[\frac{3w\pi-\pi}{3w+1}\right]-{\rm cos}\left[\frac{2\pi}{3w+1}\right]\right)^2}{(qR_{*}y^{+})^2}.
\end{equation}

\begin{figure}[htbp!]
\centering 
\includegraphics[width=0.48\textwidth]{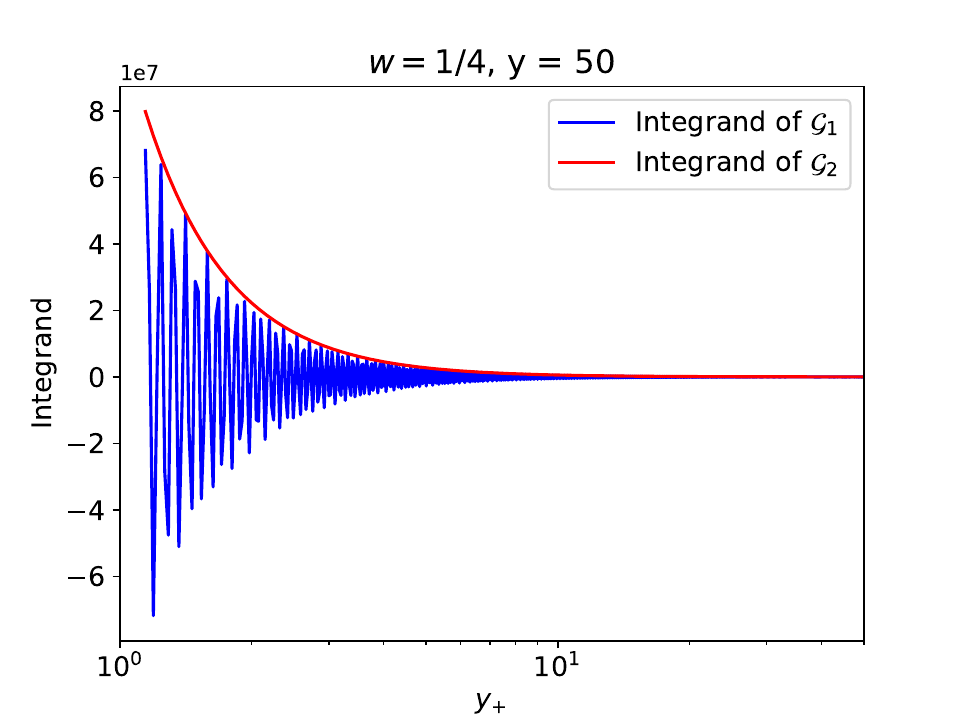}
\caption{Comparing behaviors of the integrands of $\mathcal{G}_1$ and $\mathcal{G}_2$. The blue solid line represents the integrand of $\mathcal{G}_1$, while the red solid line represents that of $\mathcal{G}_2$. 
 In both plots, $qR_{*}$ is set to 10 and $c_k$ is set to 0.00079.}
\label{fig:G1G2}
\end{figure}

Collecting all the above terms, the integral with respect to $y_{+}$ can ultimately be expressed as  
\begin{equation}
    \int {\rm d}y_{+} \frac{\mathcal{G}}{y_1^{\frac{3+3w}{2}} y_2^{\frac{3+3w}{2}}} = \mathcal{G}_1 + \mathcal{G}_2 ,
\end{equation} 
\small
\begin{align}
   \mathcal{G}_1 &= \int {\rm d}y_{+}~\bigg\{ \frac{\tilde{q}^2(1+3w)^{\frac{-4}{1+3w}}y_{+}^{\frac{-4}{1+3w}}2^{\frac{4}{1+3w}}}{({\rm cos}\left[\frac{3w\pi-\pi}{3w+1}\right]-{\rm cos}\left[\frac{2\pi}{3w+1}\right])^2} \notag\\
    &\times  \left[P_1^2\frac{{\rm cos}(2x)}{2}+ P_3^2\frac{{\rm cos}(2x')}{2}-P_1P_3{\rm cos}\left(x+x'\right) \right]\bigg\},
\end{align}
and
\begin{align} \label{eq: G2}
    \mathcal{G}_2 &= \int {\rm d}y_{+}~\bigg\{ \frac{\tilde{q}^2(1+3w)^{\frac{-4}{1+3w}}y_{+}^{\frac{-4}{1+3w}}2^{\frac{4}{1+3w}}}{\left({\rm cos}\left[\frac{3w\pi-\pi}{3w+1}\right]-{\rm cos}\left[\frac{2\pi}{3w+1}\right]\right)^2} \notag \\
    &\times {\rm cos}\left(\frac{qR_{*}y_{-}}{c_k}\right)\left[\frac{P_1^2}{2} +\frac{P_3^2}{2}-P_1P_3{\rm cos}\left(x-x'\right)\right]\bigg\}.
\end{align}
\normalsize 
Because of the cosine functions within the brackets of $\mathcal{G}_1$, the integrand curve of $\mathcal{G}_1$ oscillates rapidly, as displayed in Fig.~(\ref{fig:G1G2}) for example.
Thus, we expect its integral value just to provide the oscillatory behavior rather than the dominant term, and its integral value should be much smaller than the latter. This can be verified by using the Riemann-Lebesgue lemma if $y$ is large enough \cite{stein1993harmonic}. We also calculate its integral value over $\tilde{q}^2$ when $c_k = 0.00079$ and $qR_{*} = 10$ by numerical method and find that the corresponding value is 0.0286, -0.0035 and -0.0163 for $w = 1/4$ and $y = 5, 10 ,50$, respectively, which is much smaller than the corresponding value of $\mathcal{G}_{2}$ in most intervals. The $\mathcal{G}_2$ could be calculated directly in the limit of $y_{-} \to 0$ by noting that the term in the brackets is $y_{+}$ independent, and can be written
in the following form:
\begin{align}
     \mathcal{G}_2 &\equiv \tilde{q}^2 \frac{{\rm cos}(\tilde{q}y_{-})}{2} \Upsilon^{\nu \notin \mathbb{N}} (y;w),
\end{align}
where 
\begin{align} \label{eq: Upsilion}
 \Upsilon^{\nu \notin \mathbb{N}} &=\frac{2\left[y^{3(w-1)/2}-1\right]}{3(w-1)\left({\rm cos}\left[\frac{3w\pi-\pi}{3w+1}\right]-{\rm cos}\left[\frac{2\pi}{3w+1}\right]\right)^2} \notag \\
    &\times \left(P_1^2 +P_3^2-2P_1P_3{\rm cos}\left[\frac{3\pi(w-1)}{6w+2}\right]\right),
\end{align}
and $\mathbb{N}$ is the set of integers. By applying trigonometric identities for further simplification, we can eliminate the trigonometric functions in both the numerator and the denominator, allowing us to obtain the following final expression for $\Upsilon$ when $\nu$ is not an integer,
\begin{equation}
    \Upsilon^{\nu \notin \mathbb{N}}  =\frac{2\left[y^{3(w-1)/2}-1\right]}{3(w-1)}.
\end{equation}

\bibliography{apssamp}% Produces the bibliography via BibTeX.
\end{document}